\documentclass[pre,twocolumn,showpacs,amsmath,amssymb,noshowpacs]{revtex4}
\usepackage{amsmath,amssymb,amsthm}
\usepackage{dcolumn}
\usepackage{enumerate}
\usepackage{braket}
\usepackage{graphicx,xcolor}

\usepackage{wrapfig}
\usepackage{mathtools}
\usepackage{bm}
\usepackage{multirow}

\usepackage[%
 setpagesize=false,%
 bookmarks=true,%
 bookmarksdepth=tocdepth,%
 bookmarksnumbered=true,%
 colorlinks=true,%
 linkcolor=blue,%
 citecolor=blue,%
]{hyperref}

\usepackage[normalem]{ulem}

\usepackage{floatrow}

\theoremstyle{definition}

\theoremstyle{definition}

\definecolor{Forest}{HTML}{288c66}

\newcommand{\bv}[1]{{\boldsymbol #1}}

\makeatletter
\begin{document}
\title{Emergence of vertical diversity under disturbance

}
\author{Yuya Shishikura}
\thanks{shishikurayuya@gmail.com}
\author{Hiroki\ Ohta}\thanks{hirokiohta@obihiro.ac.jp}

\affiliation{\\
Department of Human Sciences, Obihiro University of Agriculture and Veterinary Medicine, Hokkaido, 080-8555, Japan}
\date{\today}


\begin{abstract}
We propose a statistical physics model of a neutral community, where each agent can represent identical plant species growing in the vertical direction with sunlight in the form of rich-get-richer competition. Disturbance added to this ecosystem, which makes an agent restart from the lowest growth level, is realized as a stochastic resetting.
We show that in this model for sufficiently strong competition, vertical diversity characterized by a family of Hill numbers robustly emerges as a local maximum at intermediate disturbance. 
\end{abstract}

\maketitle

\section{Introduction}
Organisms often compete with other organisms and those could coexist as a whole ecosystem. The theoretical frameworks to understand the mechanism of such coexistence in ecology have been developed based on Lotka-Volterra equation \cite{HS}. 
The developed theories have continued to overcome the obstacles to clarify unexplored coexistence mechanisms \cite{Tilman82,Chesson00,Kessler15,Obuchi16,Bunin17,Monasson17,Biroli18,Cui20,Gupta21,Barbier21,Ros23}.

One of the important topics in coexistence theory is diversity \cite{Magurran10}, which is characterized by various indexes such as Hill numbers equivalent to R{\'e}nyi entropy \cite{Leinster21}.
Diversity of an ecosystem is largely affected by so-called disturbance \cite{Viljur22}. 
Disturbance is a general term referring to the effects led by the environment outside of an ecosystem, which causes a certain loss of mass of organisms in the ecosystem. 
There is a qualitative hypothesis that high diversity is achieved at intermediate disturbance, which is called the Intermediate Disturbance Hypothesis (IDH)  \cite{Grime73, Connell78}. 
Various mathematical models have been found to show what IDH implies \cite{Petraitis89,Caswell91,Kondoh01,Roxburgh04,Svensson12,Hunt21} and also a proposal of updating such a hypothesis has been posed \cite{Fox13}. 

The concept of diversity in ecology is not only used about species, but also used about the spatial structure of an ecosystem such as the height of plants \cite{Franklin2002}.
This structural diversity has been discussed in strong relation to species diversity \cite{Inoue01,Ehbrecht21}.
Indeed, in German forests, structural diversity in terms of canopy height, which is so-called vertical diversity, has been discovered to get the highest at an intermediate disturbance \cite{Senf20}. 
Mathematical models have been also developed, which focus on height statistics of plants \cite{Hara84,Yokozawa99,Zavala07,Strigul08,Purves08,Kohyama09,Cammarano11,Kohyama12}. Nevertheless, IDH in terms of vertical diversity has hardly been mentioned from the perspective of such theoretical models.

In this paper, 
we develop a theoretical framework of statistical physics models, where one can quantitatively investigate the relationship among competition, disturbance, and vertical diversity in a community of identical species. 

The approach taken hereafter can also be regarded as a proposal on a unique branch of neutral theories in community ecology \cite{MacArthur,Caswell76,Hubbell,VH2003,MAS2004,Beeravolu09,Volkov03}. Indeed, it has already been discussed how theoretical development can go beyond neutral theories \cite{Azaele16} and how the neutral theories ignoring spatially heterogeneous effects can be extended to include such spatial effects \cite{Pigolotti18}.

\section{Model}
In this paper, we consider competition among plants for sunlight, which often takes the form of competition such that taller plants are more likely to occupy sunlight and then to get taller, which is the so-called rich-get-richer form of competition \cite{Merton68}. 

As preliminaries, suppose that for a given height distribution of plants,
we classify it into three height levels by putting two thresholds of the average plus or minus standard deviation.
Specifically, let $N\in\mathbb{Z}$ be the total number of agents, $i\in\{1,\dots,N\}$ be the name of each agent, and $x_i\in\{1,2,3\}$ be the state as the relative height of agent $i$. 
We use a notation $\bm{x}:=(x_1,x_2,\cdots,x_N)\in\{1,2,3\}^N$ as a state of the whole system.

Based on the above considerations, we introduce a stochastic process where the transition rate representing rich-get-richer competition is induced by a simple energy function $E(\bm{x})$. For simplicity, 
the energy function is required to have the following two conditions: (i) The energy is invariant in terms of replacing the state of maximum $3$ by the state of minimum $1$ in the whole system: $E(\bm{x})=E(\bm{x}')$ for $x_i'=-(x_i-2)+2$. (ii) In the configuration achieving the minimum energy, the number of state $2$ is zero and also the number of state $3$ minus the number of state $1$ is either zero or $\pm1$.

Under these two conditions, we consider the following energy function $E(\bm{x})$ in which 
each agent interacts with agents at neighboring sites determined by the edges of a graph $\mathcal{G}$,
\begin{equation}\label{energy}
    E(\bm{x}):=-\frac{1}{N-1}\sum_{i\in V(\mathcal{G})}\sum_{j\in B_i(\mathcal{G})}(x_i-x_j)^2, 
\end{equation} 
where $V(\mathcal{G})$ is the set of all sites in graph $\mathcal{G}$ and $B_i(\mathcal{G})$ for $1\le i\le N$ is the set of all the nearest neighbor sites of site $i$. 

Indeed, this energy function (\ref{energy}) satisfies (i). 
Further, (\ref{energy}) satisfies (ii) because the minimum value of the energy $-2N$ at the leading order of $N$ is realized when $x_i=1$ or $3$ for any $i$ on the condition that the number of state $1$ minus the number of state $3$ is either $0$ for even $N$ or $\pm1$ for odd $N$. 
For simplicity, we focus on only the case that $\mathcal{G}$ is the complete graph having edges by which any pair of two sites is directly connected.

Next, let us define $f_i^\pm$ such that $f^{\pm}_i\bm{x}:=(x_j\pm\delta_{i,j})^N_{j=1}$ for only the case that $1\le f_i^\pm x_i\le 3$, where $\delta_{i,j}$ is the Kronecker delta with $\delta_{a,b}=1$ for $a=b$, otherwise $0$.
Then, we introduce a Markov process in continuous time $t$ as follows. Let us consider a transition rate $T_0(\bm{x}\to f^\pm_i\bm{x})$ from $\bm{x}$ to $f^\pm_i\bm{x}$ is 
\begin{equation}\label{trate0}
    T_0(\bm{x}\to f^\pm_i\bm{x})=\frac{1}{4}\biggl[1+\tanh\bigg(-\frac{1}{2}\beta\Delta E(\bm{x}\to f^\pm_i\bm{x})\bigg)\biggr],
\end{equation} where $0\le\beta$ and $\Delta{E(\bm{x}\to f^\pm_i\bm{x})} 
        :=E(f^\pm_i\bm{x})-E(\bm{x})$
\begin{equation}\label{denergy}
        = 2\biggl\{\mp2\biggl(x_{i}-\frac{1}{N-1}\sum_{j\neq{i}}x_j\biggr)-1\biggr\}.       
\end{equation}

\begin{figure}[t]
\includegraphics[width=8cm,clip]{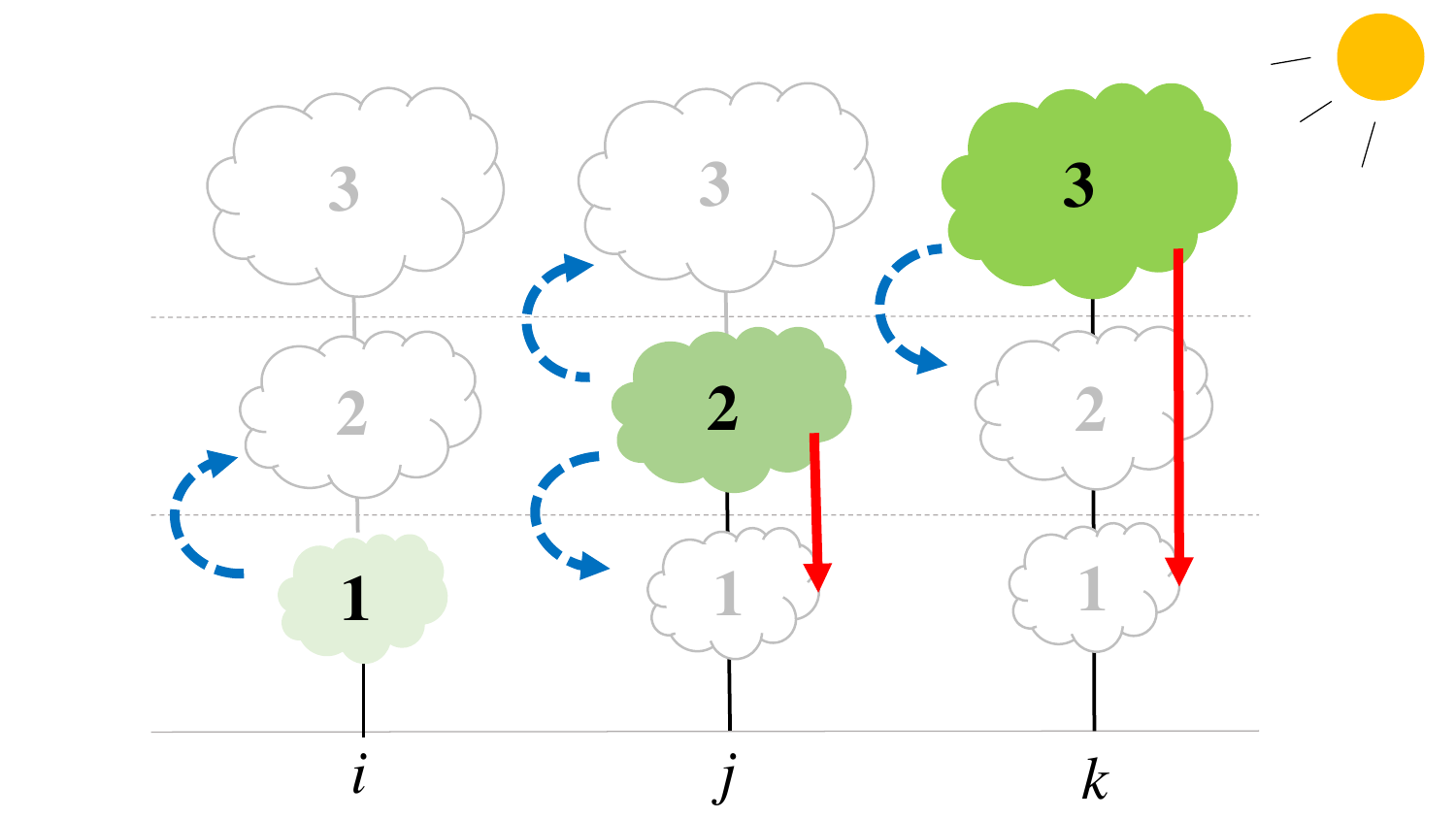}
    \caption{Schematic diagram for the transition rules between two states for the case of three plants $i, j, k$. Under sunlight competition, the transitions from $x_j=2$ to $3$ or $1$ and the reverse transitions as described by dotted blue arrows may occur. Due to disturbance, the transitions from $x_k=3$ and $x_j=2$ to $1$ described as filled red arrows may occur. } 
    \label{diag}
\end{figure}

\subsection{Stochastic resetting as disturbance}
Next, as a disturbance added to this system, we consider a transition from the current state of an agent to its lowest growth state. This transition can be regarded as one class of stochastic resetting \cite{Shamik23}. 
Concretely, let us consider stochastic resetting from state $x_i\ge 2$ to state $1$ with a disturbance rate $R(\bm{x}\to d_i\bm{x})$ where $d_i\bm{x}:=(x_j(1-\delta_{i,j}) +\delta_{i,j})^N_{j=1}$ in the following manner:
\begin{equation}\label{drate}
    R(\bm{x}\to d_i\bm{x}) = r,
\end{equation} where $0\leq r \leq 1$.

Combining this stochastic resetting with the above Markov process representing competition, the transition rate $T(\bm{x}\to \bm{x'})$ of the stochastic process that we consider, as schematically illustrated in Fig.\ \ref{diag}, is totally 
\begin{align}\label{trate}
    T(\bm{x}\to \bm{x'}) = 
    &(1-r)\sum_{1\le i\le N}\sum_{s=\pm}\delta_{\bm{x'},f^s_i\bm{x}} T_0(\bm{x}\to f^s_i\bm{x})\nonumber\\
    &+\sum_{1\le i\le N}\delta_{\bm{x'},d_i\bm{x}} R(\bm{x}\to d_i\bm{x}). 
\end{align} 
Note that the dynamics governed by the introduced rich-get-richer form of competition and disturbance is {\it neutral} in the following sense: the system has essentially the same dynamics governed by the rules of competition and disturbance even if any pairs of individuals are exchanged.

Next, let $P_t(\bm{x})$ be the probability that the state is $\bm{x}$ at time $t$. Then, transition rate (\ref{trate}) leads to the following master equation:
\begin{equation}\label{mastereq}
    \frac{d P_t(\bm{x})}{dt} = \sum_{\bm{x'}\neq\bm{x}}\Bigg(P_t(\bm{x'})T(\bm{x'}\to \bm{x})-P_t(\bm{x})T(\bm{x}\to \bm{x'})\Bigg).
\end{equation}
The stationary distribution $P_{\rm st}(\bm{x})$ is defined such that $P_t(\bm{x})=P_{\rm st}(\bm{x})$ leads to $\frac{dP_t(\bm{x})}{dt}=0$.

Indeed, the canonical distribution
$P_{\rm can}(\bm{x}):=\frac{1}{Z_N(\beta)}\exp(-\beta E(\bm{x}))$, where $Z_N(\beta)=\sum_{\bm{x}}\exp(-\beta E(\bm{x}))$, satisfies the following detailed balanced condition
\begin{equation}
P_{\rm can}(\bm{x})T_0(\bm{x}\to f^\pm_i\bm{x})=P_{\rm can}(f^\pm_i\bm{x})T_0(f^\pm_i\bm{x}\to\bm{x}).
\end{equation}
Hence, the canonical distribution is the stationary distribution satisfying Eq. (\ref{mastereq}) with $r=0$.

If the disturbance rate holds $r>0$, this stochastic process involves irreversible processes caused by disturbance, because the transition from state $1$ to state $3$ does not exist but the reverse exists. Thus, the detailed balanced condition can not be satisfied under the presence of disturbance, meaning that $P_{\rm st}(\bm{x})$ no longer follows the canonical distribution determined by the energy function (\ref{energy}).

\section{On population dynamics}

\begin{figure}[t]
   \begin{minipage}[b]{\linewidth}
\begin{flushleft}
{(a)}

\includegraphics[width=\linewidth,clip]{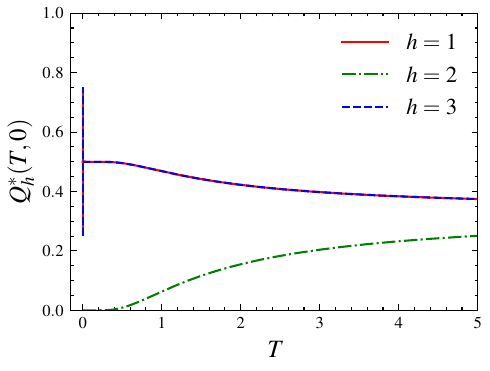} 

{(b)}
    
\includegraphics[width=\linewidth,clip]{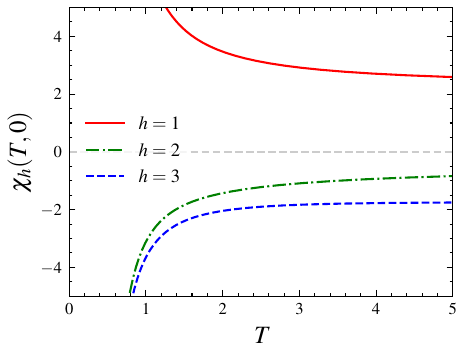}
\end{flushleft}
\end{minipage}
    \caption{(a) The density $Q_h^*(T,0)$ as a function of $T$. At $T=0$, $Q_1^*$ and $Q_3^*$ have an infinite number of solutions satisfying $1/4<Q_1^*,Q_3^*<3/4$ and $Q_2^*=0$. At $T\to\infty$, $Q_h^*(T,0)\to 1/3$ for any $h$. (b) $r$-susceptability $\chi_h(T,0)$ at $r=0$ as a function of $T$. 
    $\chi_h(T,0)$ shows an exponential divergence as $T\to 0$.}
   \label{fig:ssr0}
\end{figure}
\subsection{Derivation of population dynamics}
Although the master equation has the complete information about the stationary distribution, in order to avoid dealing with $3^N$ states, we consider the time evolution of the density of each of three different states in the limit of large $N$. We show how to derive such a set of closed equations in the following. 

First, let $Q_h$ be the density of states $h\in \{ 1, 2, 3\}$, and be $\bv{Q}:=(Q_1,Q_2,Q_3)\in[0,1]^3$. Then we focus on the transition rate $U\Bigl(h \to h\pm 1\mid \bv{Q}\Bigr)$ from state $h$ to $h\pm 1\in \{ 1, 2, 3\}$ of an agent for given $\bv{Q}$ in the limit of large $N$. Taking into account (\ref{trate0}) and (\ref{denergy}), $U$ turns out to be equal to
\begin{equation}\label{macrorate}
       \frac{(1-r)}{4}
      \biggl[1+\tanh
      \bigg(-\beta\Delta e_\pm\Big(h \mid \bv{Q}
      \Big)
      \bigg)
      \biggr],
\end{equation}
where
\begin{equation}\label{macrodenergy}
    \Delta e_\pm\Bigl(h  \mid \bv{Q}\Bigr) = \biggl\{
\mp2\biggl(h-\sum_{h'=1}^3 h' Q_{h'}\biggr)-1\biggr\}.
\end{equation} Note that it is assumed that the replacement of $\sum_{i} x_i/N$ by $\sum_{h'=1}^3 h' Q_{h'}$ in the limit of $N\to\infty$ holds, which can be regarded as self-averaging.
Taking into account disturbance rate (\ref{drate}), 
the time evolution of each density $Q_h$ for $h=1$ and $3$ is exactly described as $\frac{d Q_h}{dt} = V_h(\bv{Q})$, 
which is explicitly:
\begin{equation}\label{PDeq0}
    \begin{aligned}
      \frac{dQ_h}{dt} &=  -\delta_{h,3}rQ_3+Q_2 U\Bigl(2 \to h\mid \bv{Q}\Bigr)\\
      &+\delta_{h,1}r\sum_{h'=2}^3 Q_{h'} - Q_h U\Bigl(h \to 2 \mid \bv{Q}\Bigr).
    \end{aligned}
\end{equation}

\subsection{Densities and susceptibility in stationary states}
Before discussing vertical diversity, we try to capture the typical behaviors of the model in the parameter space of disturbance characterized by $r$ and interaction strength characterized by $T=\beta^{-1}$.

Let us compute the stationary solutions of the derived population dynamics (\ref{PDeq0}). The density of each state $h$ in a stationary solution satisfying $V_h(\bv{Q}^*)=0$ is described by $Q_h^*(T, r)$ as a function of two parameters of $T=\beta^{-1}$ and $r$. 
Note that because of $Q_1^*+Q_2^*+Q_3^*=1$, 
we have $Q_2^*=1-Q_1^*-Q_3^*$ and also $\sum_{h=1}^3hQ_h^*=-Q_1^*+Q_3^*+2$, by which Eq. (\ref{PDeq0}) is written as two coupled equations depending on only $Q_1^*$ and $Q_3^*$. 

Let us consider the case of $r=0$ for any $T$. As shown in Fig.\ \ref{fig:ssr0}(a), 
it turns out that there is a symmetric solution 
\begin{equation}
Q_1^*(T,0)=Q_3^*(T,0) = (2+\exp(-2\beta))^{-1},
\end{equation} leading to $Q_h^*(T,0)\to1/3$ for any $h$ at $T\to\infty$.
At $T=0$, 
in addition to this symmetric solution, we obtain the other asymmetric solutions as follows:
\begin{align}\label{equ:3.1.1-3}
      &\frac{1}{4}< Q_1^*(0,0)=1 - Q_3^*(0,0)<\frac{3}{4},  
\end{align} which means that  infinite number of the solutions $Q_h^*$ exist under condition (\ref{equ:3.1.1-3}).

\begin{figure}[t]
    \begin{minipage}[b]{\linewidth}
    \begin{flushleft}
    {(a)}
    
    \includegraphics[width=\linewidth,clip]{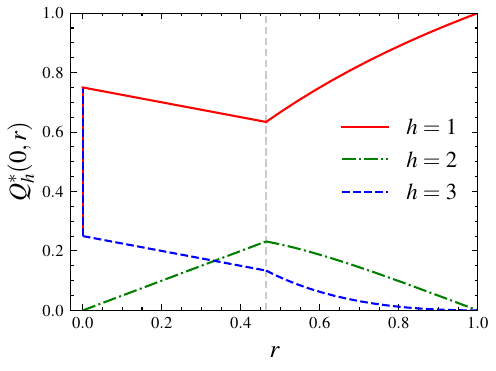}
    
    {(b)}
    
    \includegraphics[width=\linewidth,clip]{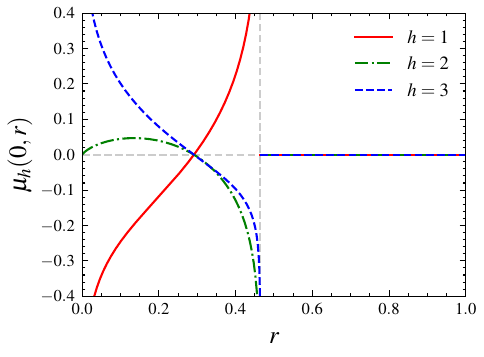}
    \end{flushleft}  
    \end{minipage}
    
    \caption{(a) The density $Q_h^*(0,r)$ as a function of disturbance rate $r$. 
    $Q_h^*$ for any $h$ are not differentiable at $r=r_0$.
    (b) $T$-susceptibility $\mu_h(0,r)$ as a function of disturbance rate $r$.
    $\mu_h(0,r)$ shows logarithmic divergences as $r\to0$ for only $h=1,3$ and $r\to r_0$ for any $h$.}
    \label{fig:ssT0}
\end{figure}

Next, let us consider a response of the density against $r$ around $r \ll 1$, which is characterized by $r$-susceptibility 
\begin{equation}
\chi_h(T,r):=\frac{\partial Q_h^*}{\partial r}
\end{equation}
for any $h$. As shown in Fig.\ \ref{fig:ssr0}(b), using Taylor series expansion, we straightforwardly obtain 
\begin{align}\label{chi0}
        &\chi_1(T,0) = 
        4\frac{(2+g_\beta)(4+g_\beta^2)+4g_\beta'\beta(2-g_\beta)}
        {(2+g_\beta)^2((2+g_\beta)(2-g_\beta)+8g_\beta'\beta)},\\
      & \chi_3(T,0) = 
      4\frac{-4g_\beta(2+g_\beta)+4g_\beta'\beta(2-g_\beta)}
      {(2+g_\beta)^2((2+g_\beta)(2-g_\beta)+8g_\beta'\beta)},
\end{align}
where $g_\beta = 1 + \tanh\beta$ with $g_0=1 \leq g_\beta < 2=\lim_{\beta\to\infty}g_\beta$ 
and $g'_\beta=(\cosh\beta)^{-2}$. 
Note that
$\chi_2=-\chi_1-\chi_3$ by definition.
It is notable that $\chi_h$ for each $h$ shows an exponential divergence as $\beta\to \infty$ ($T\to 0$).

Let us move on to the case of $T\ll 1$ for any $r>0$. 
Similar to the case with $r=0$, 
we consider a response of the density against $T$ around $T \ll 1$, which is characterized by $T$-susceptibility 
\begin{equation}
\mu_h(T,r):=\frac{\partial Q_h^*}{\partial T}
\end{equation} for any $h$.
As shown in Fig.\ \ref{fig:ssT0}, $Q_h^*$ and $\mu_h$ in the limit of $T\to0$ are computed in the following. 
First, for $r_0=-3+2\sqrt{3} < r \leq 1$, we obtain 
\begin{align}
        Q_1^*(0,r) &= \frac{2r}{1+r}, \label{Q0r1-1}\\ 
        Q_3^*(0,r) &= \frac{(1-r)^2}{(1+r)^2},\label{Q0r1-2}
\end{align} 
and $\mu_h(0,r)=0$ for any $h$.

For $0< r < r_0$, 
it is rather tricky that 
$Q_1^*(0,r)$ can not be exactly computed solely. In order to compute $Q_1^*(0,r)$, one can perform Taylor series expansion as $Q_h^*(T,r)=Q_h^*(0,r) + \mu_h(0,r)T+O(T^2)$ and then substitute it into Eq. $(\ref{PDeq0})$. 
Assuming that all the terms for each order with respect to $T$ separately satisfy Eq. $(\ref{PDeq0})$, 
we finally obtain 
\begin{align}
        & Q_1^*(0,r) = 
         Q_3^*(0,r) +\frac{1}{2}=-\frac{1}{4}r+\frac{3}{4},\label{Q0r0-1}\\
         &\mu_1(0,r)=-\frac{1+r}{1-r}\mu_3(0,r)=
        -\frac{1+r}{4}\text{artanh } \theta(r),
        \label{equ}
\end{align}
where 
\begin{equation}
    \theta(r)=\frac{-r^2-10r+3}{(1-r)(3+r)}.
\end{equation} 
    Note that $\lim_{r\to r_0}\theta(r)=-1<\theta(r)< 1=\lim_{r\to0}\theta(r)$ for $0< r < r_0$, and $\mu_2=-\mu_1-\mu_3$ by definition. 

It is notable that the average height defined by $\sum_{h=1}^3 h Q_h^*$ is equal to $3/2$ independent of $r$ for $0<r<r_0$ with $T\to0$. $\mu_h$ for $h=1,3$ as $r\to0$ and for any $h$ as $r\to r_0$ show logarithmic divergences. Further, finite size fluctuations have been found to show distinct behaviors between lower $r$ and higher $r$ compared to $r_0$ (see Appendix. B).

\begin{figure}[t]
    \includegraphics[width=\linewidth,clip]{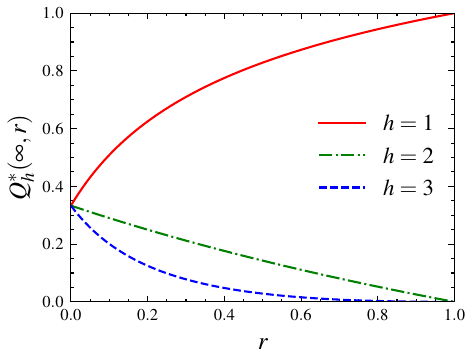}
   
    \caption{$Q_h^*$ as a function of $r$ for $\beta=0$. Note that $Q_1^* > Q_2^*,Q_3^*$  and $\chi_1>0, \chi_2<0,\chi_3<0$ hold for $0<r$.}\label{fig:solb0}
\end{figure}

For $T\to\infty$, as shown in Fig.\ \ref{fig:solb0}, we can straightforwardly obtain the analytical form of $Q_h^*(\infty,r)$ and $\chi_h(\infty,r)$ as 
\begin{align}
        Q_1^*(\infty,r) = \frac{5r^2+10r+1}{3r^2+10r+3},\label{0solb0-1}\\
        Q_3^*(\infty,r) = \frac{(1-r)^2}{3r^2+10r+3},\label{0solb0-2}\\
         Q_2^*(\infty,r) = \frac{-3r^2+2r+1}{3r^2+10r+3},\label{0solb0-3}
\end{align}
which satisfy $Q_1^*>Q_2^*,Q_3^*$ and $\chi_1>0, \chi_2<0, \chi_3<0$ for any $r>0$ as shown in Fig.\ \ref{fig:solb0}. Note that $\nu_h(\infty,r):=\frac{\partial Q_h^*}{\partial \beta}=0$ holds for any $h$.

\section{Diversity indexes}
Let us focus on Hill numbers for $\alpha>0$ and $\alpha\neq 1$,
\begin{equation}
       D_\alpha :=  \biggl(\sum_{h=1}^3 Q_{h}^\alpha \biggl)^{1/(1-\alpha)},
\end{equation} 
which satisfies the replication principle \cite{Leinster21} and quantifies the vertical diversity of the model with $1\le D_\alpha\le 3$. 
Further, we define 
\begin{equation}
    D_1:=\lim_{\alpha\to1} D_\alpha=\exp(-\sum^3_{h=1}Q_h^*\log_e{Q_h^*}),
\end{equation} 
which is equivalent to the exponential of Shannon entropy, where $e$ is Napier's constant.

\begin{figure}[t]
    \begin{minipage}[b]{\linewidth}
    \begin{flushleft}
    (a)

    \includegraphics[clip, width=\linewidth]{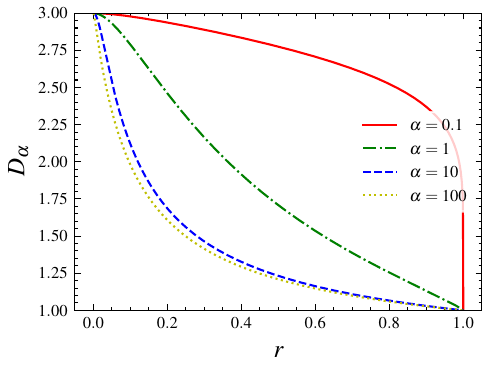}

 (b)
 \includegraphics[width=\linewidth,clip]{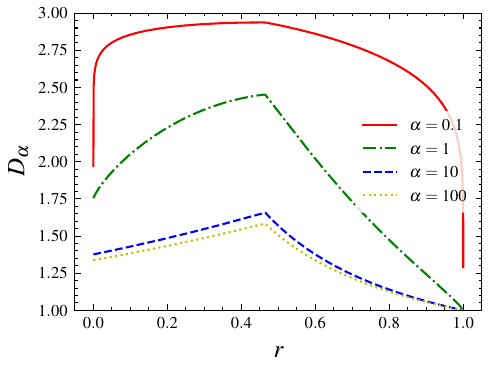}
    \label{divT0(a)}
    
    \end{flushleft}
    \end{minipage}    
    \caption{
    Diversity indexes $D_\alpha(r)$ as a function of $r$ for $T\to\infty$ (a) 
    and $T\to0$. (b)  $\alpha=0.1, 1, 10, 100$.
    $D_\alpha(r)$ do not have local maxima as a function of $r$ at $T\to\infty$. On the contrary, at $T\to0$, $D_\alpha(r)$ have local maxima at intermediate values of disturbance $r$ for any $\alpha$.
     }\label{fig:diversity}
\end{figure}

In Fig.\ \ref{fig:diversity}, $D_\alpha$ for $\alpha=0.1,1,10, 100$ are shown at $T\to 0$ and $T\to \infty$. Whereas the diversity indexes at $T\to \infty$ show monotonically decreasing behaviors, 
those at $T\to 0$ robustly get a maximum 
as a function of a single 
variable $r$ at an intermediate disturbance $r=r_0$ for $\alpha \ge 0.199$.

In order to elaborate on Hill numbers, we discuss theoretical results in the cases of $T\to\infty$ and $T\to0$ in the following subsection A. Then, in the following subsection B, we present numerical computation of Hill numbers for more general values of $T$.

\subsection{Hill numbers $D_\alpha(r)$ as a function of $r$}

In order to discuss how the diversity indexes behave for general values of $\alpha$,
let us define $M_\alpha:=\log D_\alpha$ and $\bv{Q}^*:=(Q_1^*,Q_2^*,Q_3^*)\in[0,1]^3$. 
Then, the sign of $\partial_r M_\alpha$ is the same as that of $\partial_r D_\alpha$. 
Indeed, we get the following:
\begin{align} \label{difflogH}
    &\partial_r M_\alpha
    =\alpha(1-\alpha)^{-1}
    (\sum_h (Q_h^*)^\alpha)^{-1}F_\alpha(\bv{Q}^*),\\   
    &F_\alpha(\bv{Q}^*)
    =\sum_h (Q_h^*)^{\alpha-1}\chi_h.
\end{align}
Thus, the nontrivial part determining the sign of $\partial_r D_\alpha$ is $F_\alpha$, which has key information to find out the local maximum point $r_\ell(T,\alpha)$ of $D_\alpha$. Note that 
$F_\alpha=0$ at $\alpha=1$ holds due to $\sum_h \chi_h=0$.

\subsubsection{The absence of local maxima of $D_\alpha$ at 
$T\to\infty$ 
for $0<\alpha<\infty$}

Suppose the following condition 
that for a fixed $T$, 
 $Q_1^*>Q_2^*, Q_3^*$ and $\chi_1>0$, $\chi_2<0$, $\chi_3<0$ hold for $0<r\le 1$. On that condition, we can derive that $(Q_1^*)^{\alpha-1}\chi_1$ as a term in $F_\alpha$ can be larger than $|(Q_2^*)^{\alpha-1}\chi_2+(Q_3^*)^{\alpha-1}\chi_3|$ for $\alpha>1$, 
and smaller than it for $0<\alpha<1$. This leads to that for $0<r\le 1$, $F_\alpha >0$ holds for $\alpha>1$, 
and $F_\alpha<0$ holds for $0<\alpha<1$, corresponding to that there is no local maximum of $D_\alpha$ as a function of $r$ in $0<r\le 1$ for $\alpha>0$. 

\begin{figure}[t]
    \begin{minipage}[b]  
    {\linewidth}
    \begin{flushleft}
    {(a)}
    
    \includegraphics[width=1\linewidth,clip]{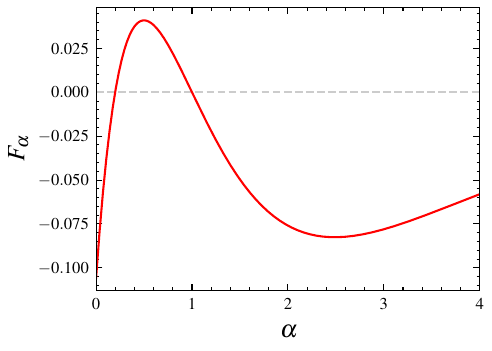}

    {(b)}

    \includegraphics[width=\linewidth,clip]{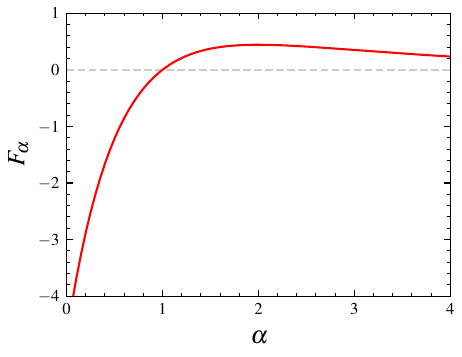}
    
    \end{flushleft}
     \end{minipage} 
    
    \caption{(a) $\lim_{r\to r_0}F_\alpha$ 
    as a function of $\alpha$ for $r<r_0$ and (b) for $r>r_0$. Reminding of (\ref{difflogH}), it turns out that $\partial_ r\log D_\alpha$ are negative for any $\alpha$ in the case of $r>r_0$ and for, at least, $\alpha\ge 0.199$ in the case of $r<r_0$. On the contrary, $\partial_ r\log D_\alpha$ are positive for, at least, $\alpha\le 0.198$ in the case of $r<r_0$, which means that the local maxima for $\alpha\le 0.198$ are not located at $r=r_0$.}
    \label{fig:F_alpha_r0}
\end{figure}

The derivation of the above conclusion is as follows. First, it is straightforward to see
\begin{align}
&|(Q_2^*)^{\alpha-1}\chi_2+(Q_3^*)^{\alpha-1}\chi_3|\nonumber
\\
&= 
(Q_2^*)^{\alpha-1}\chi_1
+(Q_3^*)^{\alpha-1}|\chi_3|
-(Q_2^*)^{\alpha-1}|\chi_3|\nonumber\\
&=(Q_3^*)^{\alpha-1}\chi_1
+(Q_2^*)^{\alpha-1}|\chi_2|
-(Q_3^*)^{\alpha-1}|\chi_2|
,\label{Lineq1}
\end{align}
where we have used $\sum_h\chi_h=0$. Hence, we obtain
\begin{align}
&\chi_1\min(Q_2^*,Q_3^*)^{\alpha-1}\nonumber\\
&\le 
|(Q_2^*)^{\alpha-1}\chi_2+(Q_3^*)^{\alpha-1}\chi_3|\nonumber\\
&\le 
\chi_1\max(Q_2^*,Q_3^*)^{\alpha-1}
.\label{Lineq2}
\end{align}
It means that for $\alpha>1$, 
the following holds:
\begin{align}
\chi_1(Q_1^*)^{\alpha-1}>\chi_1\max(Q_2^*,Q_3^*)^{\alpha-1},
\end{align} which 
immediately leads to $F_\alpha > 0$.
On the other hand, 
for $0<\alpha<1$, 
the following holds:
\begin{align}
\chi_1(Q_1^*)^{\alpha-1}<\chi_1\min(Q_2^*,Q_3^*)^{\alpha-1},
\end{align} 
which immediately leads to 
$F_\alpha < 0$. 
Thus, we reach the conclusion mentioned above. 

Note that for $T=\infty$, (\ref{0solb0-1}),(\ref{0solb0-2}), and (\ref{0solb0-3}) directly mean that the conditions of $Q_1^*>Q_2^*,Q_3^*$ and $\chi_1>0, \chi_2<0, \chi_3<0$ hold.
Thus, it leads to the absence of local maxima of $D_\alpha$ in $0<r\le1$ for any $\alpha>0$ as shown in Fig.\ \ref{fig:diversity}(a).

\subsubsection{The existence of a local maximum of $D_\alpha$ at $T\to0$ for $0<\alpha<\infty$}

First, let us consider the case of $r<r_0$ and $T\to0$ where we use Eq. (\ref{Q0r0-1}) by putting $r=r_0$. Then, as shown in Fig.\ \ref{fig:F_alpha_r0}(a), we obtain that $F_\alpha <0$ for $\alpha>1$, $F_\alpha>0$ for $0.199\le\alpha<1$, 
and $F_\alpha<0$ for $\alpha\le 0.198$ as $r\to r_0$. 
It means that for $\alpha \ge 0.199$, 
$\partial_r D_\alpha$ as $r\to r_0$ is positive, whereas it is negative for $0<\alpha \le 0.198$.

Second, let us consider the case of $r>r_0$ 
where we use Eqs. (\ref{Q0r1-1}) and (\ref{Q0r1-2})
by putting $r=r_0$. Then, as shown in Fig.\ \ref{fig:F_alpha_r0}(b), 
it turns out that $F_\alpha>0$ for $\alpha>1$ and $F_\alpha<0$ for $\alpha<1$. It means that $\partial_r D_\alpha$  as $r\to r_0$ is negative for any $\alpha>0$.

Therefore, combining two cases together, we reach the conclusion that there is a local maximum point $r_\ell(0,\alpha)>0$ at $T\to0$ for any $\alpha>0$.
The concrete values of $r_\ell(0,\alpha)$ are $r_0$ for $\alpha\ge0.199$, and get smaller than $r_0$ between $\alpha=0.198$ and $0.199$. Then, $\lim_{\alpha\to0}r_\ell(0,\alpha)$ is numerically estimated as $0.451\cdots$.

\begin{figure}[t] 
    \includegraphics[width=\linewidth,clip]{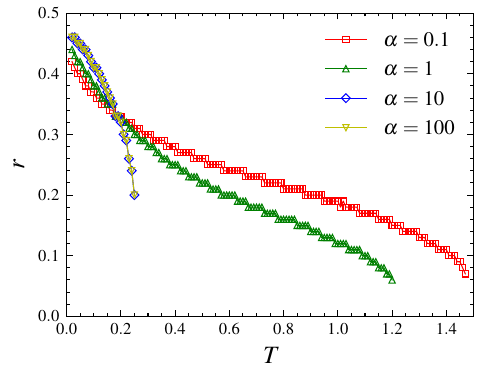}\label{div_phasechart(b)}
    \caption{
    The local maximum points $r_{\ell}(T,\alpha)$ of $D_\alpha(r)$ depending on $(T,\alpha)$ for $Q_h^*$. $r_{\ell}(0,0.1)=0.457\cdots< r_0=0.464\cdots$.
    The data points of $r_{\ell}(T,\alpha)$ are estimated within 
    the numerical resolution by a nullcline method where the increment of $r$ and $T$ is $10^{-2}$. $T=0.25$ is around the value, above which local maxima for certain values of $\alpha$ do not appear.}
    \label{diversity0}
\end{figure}

\subsection{Computation of Hill numbers by nullcline method}

Next, we have performed the nullcline method for computing $\bv{Q}^*$ to estimate the local maximum of $D_\alpha(r)$ as a function of $r$ for general values of $T$ and $\alpha$.
As shown on the $(T,r)$-parameter plane 
in Fig.\ \ref{diversity0}, 
whereas $r=0$ could also get the local or global maximum point for any $T>0$, local maximum points $r_{\ell}(T,\alpha)>0$ of $D_\alpha$ as a function of single variable $r$ are determined by $\partial_rD_\alpha(r_{\ell}(T,\alpha))=0$. 
Indeed, there is a point around $T=0.25$ below which there exists a local maximum point for any $\alpha>0$.
On the contrary, for $T\ge 0.26$,  $r_{\ell}(T,\alpha)$ does not exist for all the values of $\alpha$ higher than a sufficiently large $\alpha_{0}(T)$.

Related to this phenomenon, 
we have also found that the local minima of 
$Q_1^*$ as a function of single variable $r$, 
which are observed for $T\le 0.25$, 
disappear for $T\ge0.26$.
Further, the local maxima of $Q_2$ exist for $T\le2.58$, and it seems to disappear between $T=2.58$ and $T=2.60$, which is very close to the lower limit of the absence of local maxima of $D_\alpha$ for any $\alpha$.

\section{Concluding remarks}
In this paper, 
we have proposed a neutral community model of identical plant species with rich-get-richer competition, which robustly exhibit emergent vertical diversity that has a local maximum at an intermediate disturbance for sufficient strong competition. 
This property observed in the proposed model is qualitatively consistent with the property observed in German forests \cite{Senf20}. 
It remains for future studies to perform more detailed comparisons with the properties observed in real ecosystems.

It is intriguing to consider how to blend the effects of migration and death process from classical neutral models \cite{Hubbell,VH2003,MAS2004} with the model in this paper. Future works in that direction could provide unique insights into neutral species abundance distributions into vertically-structured local communities.

The proposed neutral model can be easily generalized into a non-neutral model with multiple species 
by introducing species dependence into transition rates representing competition or disturbance. In another direction, one can study spatial effects by considering the model on a finite-dimensional lattice 
such as a one-dimensional chain instead of the complete graph. Based on finite-dimensional lattice models, a mixing process could also be introduced as a parameter to study how such lattice models go through a transition to the corresponding population dynamics such as a Lotka-Volterra type of equation \cite{FMO}. The relationship between lattice models and population dynamics is one topic discussed in the context of intermediate disturbance hypothesis \cite{Svensson12}.
Those cases remain to be studied.

\section*{Acknowledgements}
The authors thank Obihiro University of Agriculture and Veterinary Medicine for providing the facilities to accomplish this work.




\appendix

\section{The equations to determine the stationary solutions of the population dynamics}\label{App:}

As mentioned in the main text, based on the model we study, we consider $\bv{Q}:=(Q_1,Q_2,Q_3)\in[0,1]^3$ where $Q_h$ is the density of each state $h$. Then, we may derive the following equation of the population dynamics of $\bv{Q}$ for $h=1,3$: 
    \begin{equation}\label{PDeq}
    \begin{aligned}
      &\frac{d Q_h}{dt} = V_h(\bv{Q}),\\
      &V_h(\bv{Q}) =  -\delta_{h,3}rQ_3+Q_2 U\Bigl(2 \to h\mid \bv{Q}\Bigr)\\
      &+\delta_{h,1}r\sum_{h'=2}^3 Q_{h'} - Q_h U\Bigl(h \to 2 \mid \bv{Q}\Bigr).
    \end{aligned}
\end{equation} 
We define $Q_h^*$ such that $V_h(\bv{Q}^*)=0$ holds for $h=1,3$.

Let us remember that because of $Q_1^*+Q_2^*+Q_3^*=1$, we have
\begin{align}
&Q_2^*=1-Q_1^*-Q_3^*,\label{Q2}\\
&\sum_{h=1}^3hQ_h^*=-Q_1^*+Q_3^*+2.\label{SumQ}
\end{align} 
We substitute (\ref{Q2}) and (\ref{SumQ}) into (\ref{PDeq}) and then obtain the following for $h=1,3$:
\begin{equation}\label{PDst}
    \begin{aligned}
      & - \delta_{h,3}rQ_3^*+(1-Q_1^*-Q_3^*)\frac{(1-r)}{4}\Big(1+G_\beta^{(h)}(Q_3^*-Q_1^*)\Big) \\
      &+ \delta_{h,1}r(1-Q_1^*)  - Q_h^*\frac{(1-r)}{4}\Big(1-G_\beta^{(h)}(Q_3^*-Q_1^*)\Big) = 0,
    \end{aligned}
\end{equation}
where 
\begin{equation}
G_\beta^{(h)}(a) = \tanh\beta\Big( (\delta_{h,1}-\delta_{h,3})2a+1\Big).
\end{equation}

\subsection{The explicit expressions of $Q_h^*$ and $\chi_h$ at $r\to0$}

Based on (\ref{PDst}) for $T>0$, we find the stationary solution of the densities $Q_1^*$ and $Q_3^*$ satisfying the following:
\begin{equation}
    \begin{aligned}
      & \frac{1}{4}(1-2Q_1^*)(1+\tanh\beta) - \frac{1}{4}Q_1^*(1-\tanh\beta) = 0.\\ 
    \end{aligned}
\end{equation}
Then, taking into account (\ref{Q2}), we obtain 
\begin{align}
      & Q_1^*(T,0) = Q_3^*(T,0) = \frac{1}{2+\exp(-2\beta)}, \label{Tsol1}\\ 
      & Q_2^*(T,0) = \frac{\exp(-2\beta)}{2+\exp(-2\beta)}.\label{Tsol2}
\end{align}

Next, we focus on the case of $T=0$ where the following holds: 
\begin{equation}\label{Gvalues}
    \lim_{T\to 0} G_\beta^{(h)}(a) = 
    \begin{cases}
    1 & {\rm if}\ \beta\bigg((\delta_{h,1}-\delta_{h,3})2a+1\bigg) > 0\\
    -1 & {\rm if}\ \beta\bigg((\delta_{h,1}-\delta_{h,3})2a+1\bigg) < 0\\
    0 & {\rm if}\ \beta\bigg((\delta_{h,1}-\delta_{h,3})2a+1\bigg) =0.
    \end{cases}
\end{equation}
There are six cases in terms of the possibilities of taking values of $G_\beta^{(1)}(a)$ and $G_\beta^{(3)}(a)$, each of which can be $-1$, $0$, or $1$. Note that if one of $G_\beta^{(1)}(a)$ and $G_\beta^{(3)}(a)$ is $0$, 
the value of the other is uniquely determined.
Indeed, when the two conditions of 
\begin{align}
    -2Q_1^*+2Q_3^*+1>0, \label{T0cond1-1}\\
    2Q_1^*-2Q_3^*+1>0, \label{T0cond1-2}
\end{align} hold, based on (\ref{PDst}), we obtain
\begin{equation}
    \begin{aligned}
      Q_3^* = -Q_1^* + 1,
    \end{aligned}
\end{equation}
where we have used 
\begin{equation}
    \lim_{T\to0}G_\beta^{(h)}(Q_3^*-Q_1^*)=1,
\end{equation} for $h=1,3$.
Additionally, the two conditions of (\ref{T0cond1-1}) and (\ref{T0cond1-2}) lead to

\begin{align}
      & \frac{1}{4}< Q_1^*(0,0) <\frac{3}{4}, \label{T0sol1} \\
      & Q_3^*(0,0) = 1 - Q_1^*(0,0),  \label{T0sol2}\\
      & Q_2^*(0,0) = 0. \label{T0sol3}
\end{align}
Note that the other five cases for the possibilities of taking values of $G_\beta^{(h)}(a)$ give no solutions of Eq. (\ref{PDst}).

In order to compute $\chi_h(T,r):=\frac{\partial Q_h^*}{\partial r}$ for $T>0$, 
let us perform Taylor series expansion of $Q_h^*(T,r)$ at $r=0$ for $h=1, 3$ as follows:
\begin{equation}
    \begin{aligned}
        Q_h^*(T,r) = Q_h^*(T,0) + \chi_h(T,0)r + O(r^2).  
        \label{equ:A.2-1}
    \end{aligned}
\end{equation}
Then, substituting this form with (\ref{Tsol1}) and (\ref{Tsol2}) into (\ref{PDst}), for $O(\beta r)\ll 1$ up to $O(r)$ order, we obtain  
\begin{equation}
    \begin{aligned}
          & -2\chi_1 - g_\beta\chi_3 + \frac{4g_\beta'\beta}{2+g_\beta}(\chi_3-\chi_1)+ \frac{8}{2+g_\beta} = 0, \\
          & 
          -g_\beta\chi_1 - 2\chi_3 + \frac{4g_\beta'\beta}{2+g_\beta}(\chi_1-\chi_3)
          - \frac{4g_\beta}{2+g_\beta} = 0,
        \label{equ:3.1.4-5}
    \end{aligned}
\end{equation}
where
\begin{align}
        &g_\beta = 1 + \tanh\beta,\\
        &g_\beta'=\frac{1}{\cosh^2\beta},
\end{align} with $g_0=1 \leq g_\beta < \lim_{\beta\to\infty} g_\beta=2$.

Solving these equations, 
we finally obtain 
\begin{align}
        &\chi_1(T,0) = 
        4\frac{(2+g_\beta)(4+g_\beta^2)+4g_\beta'\beta(2-g_\beta)}
        {(2+g_\beta)^2((2+g_\beta)(2-g_\beta)+8g_\beta'\beta)}, \label{chi0A1}\\
      & \chi_3(T,0) = 
      4\frac{-4g_\beta(2+g_\beta)+4g_\beta'\beta(2-g_\beta)}
      {(2+g_\beta)^2((2+g_\beta)(2-g_\beta)+8g_\beta'\beta)}, \label{chi0A2}
\end{align}
where $g_\beta'\beta \simeq \beta\exp(-2\beta)\to 0$ as $\beta\to\infty$ and $g_\beta'\beta\simeq \beta\to 0$ as $\beta\to0$.
Note that $\chi_2=-\chi_1-\chi_3$ by definition.

\subsection{The explicit expressions of $Q_h^*$ and $\mu_h$ at $T\to0$}

Among six possibilities of taking values of $G_\beta^{(1)}$ and $G_\beta^{(3)}$, 
let us consider the situation where the following two conditions hold for $T\ll 1$:  
\begin{align}
-2Q_1^*+2Q_3^*+1=0, \label{T0cond2-1} \\ 
2Q_1^*-2Q_3^*+1 > 0. \label{T0cond2-2}
\end{align}  
In this case, 
we need to take the limit of $T\to0$ carefully. 
First, we can perform 
Taylor series expansion of $Q_h^*(T,r)$ at $T=0$ as
\begin{equation}
    \begin{aligned}
        Q_h^*(T,r) = Z_h + \mu_h(0,r)T + O(T^2),
    \end{aligned}
\end{equation} where $Z_h:=Q_h^*(0,r)$ and $\mu_h(T,r):=\frac{\partial Q_h^*}{\partial T}$.
Substituting this form into (\ref{PDst}), 
we obtain, up to $O(T^0)$ order,  
\begin{align}
      &\frac{(1-r)}{4}(1-Z_1-Z_3)\Bigl\{1+\tanh2(-\mu_1+\mu_3) \Bigr\} \nonumber\\
        &- \frac{(1-r)}{4}Z_1\Bigl\{1-\tanh2(-\mu_1+\mu_3) \Bigr\} + r(1-Z_1) = 0,\label{T0eq1}\\ 
         &\frac{1}{2}(1-r)(1-Z_1-Z_3) - rZ_3 = 0,\label{T0eq2}
\end{align}
where the second equation is obtained on the condition of (\ref{T0cond2-1}) 
corresponding to
\begin{align}\label{Glimit}
 \lim_{T\to0}G_\beta^{(3)}(Q_3^*-Q_1^*)\to 1.
\end{align}

Combining (\ref{T0eq2}) with the condition of (\ref{T0cond2-1}) gives

\begin{align}
       Z_1 = & Q_1^*(0,r) = -\frac{1}{4}r+\frac{3}{4}, \label{T0rsol1}\\ 
       Z_3 = & Q_3^*(0,r) = -\frac{1}{4}r+\frac{1}{4}, \label{T0rsol2}\\
       &Q_2^*(0,r) = \frac{r}{2}. \label{T0rsol3}
\end{align}

Let us move onto (\ref{T0eq1}) into which we substitute (\ref{T0rsol1}) and (\ref{T0rsol2}),
leading to
\begin{equation}\label{keyrelation}
    \begin{aligned}
      \tanh2(-\mu_1+\mu_3)  &= \frac{-r^2-10r+3}{(r+3)(1-r)}=:\theta(r).
    \end{aligned}
\end{equation}
Since $-1 < \tanh2(-\mu_1+\mu_3)< 1$ should hold, $r$ satisfies $0 < r < r_0=-3+2\sqrt{3}$ due to
$\lim_{r\to 0}\theta(r)=1$ and $\lim_{r\to r_0}\theta(r)=-1$. 
Thus, it turns out that (\ref{T0rsol1}), (\ref{T0rsol2}) and (\ref{T0rsol3}) as the stationary solution is valid only for $0 < r < r_0$.
Note that $\lim_{r\to0} Q_1^*(0,r)=\sup Q_{1}^*(0,0)=\dfrac{3}{4}$, $\lim_{r\to0} Q_3^*(0,r)=\inf Q_{3}^*(0,0)=\dfrac{1}{4}$, 
and $\lim_{r\to0} Q_2^*(0,r)=Q_{2}^*(0,0)=0$
hold.

Next, among the other five possibilities of taking values of $G_\beta^{(1)}$ and $G_\beta^{(3)}$, when the following two conditions 
\begin{align}
    -2Q_1^*+2Q_3^*+1<0, \label{T0cond3-1}\\
    2Q_1^*-2Q_3^*+1>0, \label{T0cond3-2}
\end{align} 
hold, based on  (\ref{PDst}) we obtain
\begin{align}
        -\frac{1}{2}(1-r)Q_1^* + r(1-Q_1^*) = 0, \\ 
        \frac{1}{2}(1-r)(1-Q_1^*-Q_3^*) - rQ_3^* = 0,
\end{align}
where we have used 
\begin{align}
\lim_{T\to0}G_\beta^{(1)}(Q_3^*-Q_1^*)=-1,\\
\lim_{T\to0}G_\beta^{(3)}(Q_3^*-Q_1^*)=1.
\end{align}
It reads 
\begin{align}
      Q_1^*(0,r) = \frac{2r}{1+r}, \label{T0rsol2-1}\\
      Q_3^*(0,r) = \frac{(1-r)^2}{(1+r)^2}, \label{T0rsol2-2}\\
      Q_2^*(0,r) = \frac{2r(1-r)}{(1+r)^2}. \label{T0rsol2-3}
\end{align}
Substituting (\ref{T0rsol2-1}) and (\ref{T0rsol2-2}) into 
the two conditions of (\ref{T0cond3-1}) and (\ref{T0cond3-2}) 
gives
\begin{align}
  3r^2+10r-1>0, \\ 
  r^2+6r-3>0, 
\end{align} 
which lead to $r_0< r \le 1$. 
Thus, the obtained solution of (\ref{T0rsol2-1}), (\ref{T0rsol2-2}), and (\ref{T0rsol2-3}) is valid only for $r_0<r\le 1$.
Combining two cases of smaller $r$ and larger $r$ than $r_0$, the left-sided limit and the right-sided limit of $Q_h^*$ as $r\to r_0$ coincide.
Note that the other four cases for the possibilities of taking values of $G_\beta^{(h)}(a)$ give no solutions of Eq. (\ref{PDst}).

In order to compute the response $ \mu_h(0,r)$, we can perform Taylor series expansion as 
\begin{equation}
    \begin{aligned}
        Q_h^*(T,r) = Q_h^*(0,r) + \mu_h(0,r)T + O(T^2).
    \end{aligned}
\end{equation}
Substituting this form into (\ref{PDst}) with (\ref{Glimit}) for $0<r<r_0$, we
obtain, up to $O(T)$ order,
\begin{align}
&\frac{(1-r)}{4}(1-Q_1^*-Q_3^*)\Bigl\{1+\tanh2(-\mu_1+\mu_3+O(T)) \Bigr\} \nonumber\\
        &- \frac{(1-r)}{4}Q_1^*\Bigl\{1-\tanh2(-\mu_1+\mu_3
        +O(T)) \Bigr\} + r(1-Q_1^*) = 0,\label{mu0eq}
        \\ 
         &(1-r)\mu_1+(1+r)\mu_3=0.\label{mueq}
\end{align}
Note that the first equation does not give an explicit exact solution up to $O(T)$ order.

Combining (\ref{mueq}) with (\ref{keyrelation}), we obtain
\begin{align}
        \mu_1(0,r) = -\frac{1+r}{4}\text{artanh }\theta(r),\\ \mu_3(0,r) = \frac{1-r}{4}\text{artanh }\theta(r).
\end{align}
Note that $\mu_2 = -\mu_1 -\mu_3$ holds by definition, leading to 
\begin{equation}
    \begin{aligned}
        \mu_2(0,r) = \frac{r}{2}\text{artanh }\theta(r).
    \end{aligned}
\end{equation}
Thus, it turns out that $|\mu_h|\simeq \log (r-r_0)$ for each $h$.

Let us move onto the case of  $r_0 < r \le 1$. By using the same Taylor series expansion, instead of (\ref{mu0eq}), we obtain
\begin{equation}
        -r\mu_1 - \frac{1}{2}(1-r)\mu_1 = 0.
\end{equation}
Then, combining this equation with (\ref{mueq}) leads to
\begin{equation}
    \begin{aligned}
         \mu_1(0,r) = \mu_3(0,r) = \mu_2(0,r)=0.
    \end{aligned}
\end{equation}

\subsection{The explicit expressions of $Q_h^*$ and $\nu_h$ at $\beta\to0$}
Based on (\ref{PDst}), taking into account 
\begin{equation}
    \lim_{\beta\to 0} G_\beta^{(h)}(a) = 0,
\end{equation}
we obtain the following:
\begin{align}
        \frac{(1-r)}{4}(1-Q_1^*-Q_3^*) - \frac{(1-r)}{4}Q_1^* + r(1-Q_1^*) &= 0,\\ 
        \frac{(1-r)}{4}(1-Q_1^*-Q_3^*) - \frac{(1-r)}{4}Q_3^* - rQ_3^* &= 0.
\end{align}
Solving these equations, we obtain
\begin{align}
        Q_1^*(\infty,r) = \frac{5r^2+10r+1}{3r^2+10r+3}, \label{solb0-1}\\
        Q_3^*(\infty,r) = \frac{(1-r)^2}{3r^2+10r+3}, \label{solb0-2}\\
         Q_2^*(\infty,r) = \frac{-3r^2+2r+1}{3r^2+10r+3}.\label{solb0-3}
\end{align}

In order to compute the response $\nu_h(T,r) := \frac{\partial Q_h^*}{\partial \beta}$, we perform Taylor series expansion of $Q_h^*(T,r)$ at $T=\infty$ as
\begin{equation}
    Q_h^*(T,r) = Q_h^*(\infty,r) + \nu_h(\infty,r)\beta + O(\beta^2).
\end{equation}
Substituting this form into (\ref{PDst}), we obtain
    \begin{align}
        -\frac{1-r}{4}(\nu_1+\nu_3) - \frac{1-r}{4}\nu_1 - r\nu_1 &= 0,\\ 
        -\frac{1-r}{4}(\nu_1+\nu_3) - \frac{1-r}{4}\nu_3 - r\nu_3 &= 0.
    \end{align}
Solving these equations, we obtain
\begin{equation}
        \nu_1 = \nu_3 = 0,
\end{equation} leading to  $\nu_2 = 0$ by definition. 

Note that by taking the derivative of the stationary solutions from (\ref{solb0-1}), (\ref{solb0-2}), and (\ref{solb0-3}) with respect to $r$, we also have
\begin{align}
\chi_1(\infty,0)=\frac{20}{9},\\
\chi_3(\infty,0)=-\frac{16}{9},\\
\chi_2(\infty,0)=-\frac{4}{9},
\end{align} which 
are consistent with the expression of (\ref{chi0A1}) and (\ref{chi0A2}) in the limit of $\beta\to0$.

\section{Monte Carlo sampling}

\begin{figure}[t]
    \includegraphics[width=\linewidth,clip]{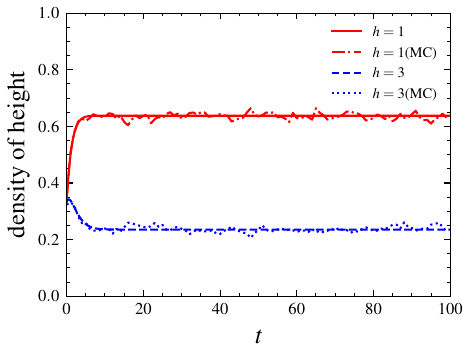}
    \caption{The density of height, $\sum_i\delta_{x_i,h}/N$, calculated by the Monte Carlo (MC) simulation with $N=1000$ and the solutions $Q_h$ of the population dynamics (\ref{PDeq}). 
    $T=1$, $r=0.2$.} \label{dynamicsdata}
\end{figure}

\subsection{Updating rules}

\begin{figure*}[t]
    \begin{center}   \includegraphics[scale=0.7]{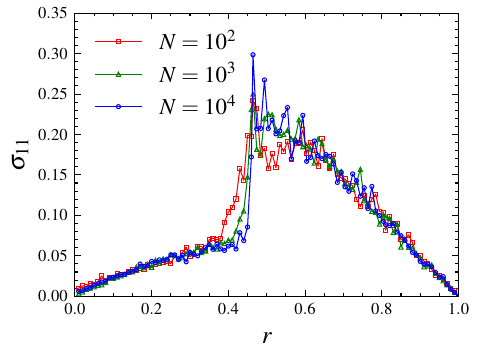}\includegraphics[scale=0.7]{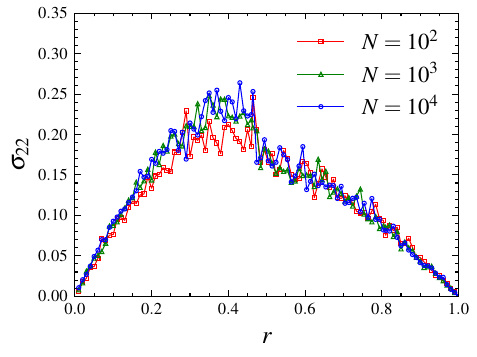}\includegraphics[scale=0.7]{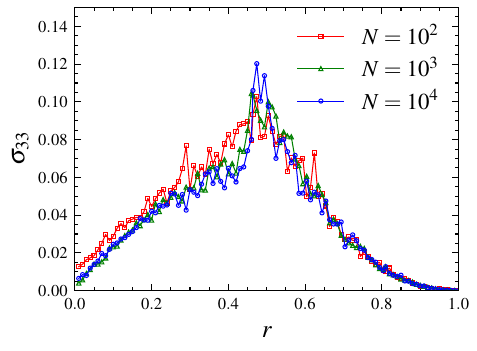} \\ 
    \includegraphics[scale=0.7]{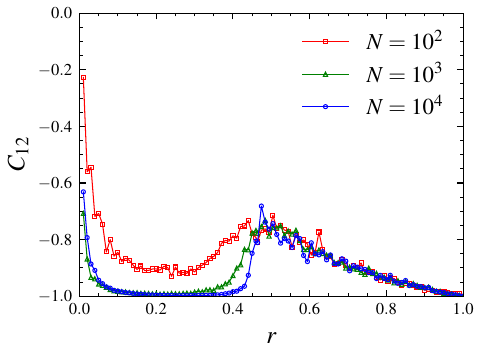}\includegraphics[scale=0.7]{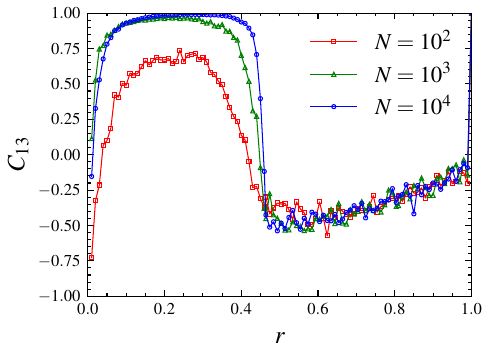}\includegraphics[scale=0.7]{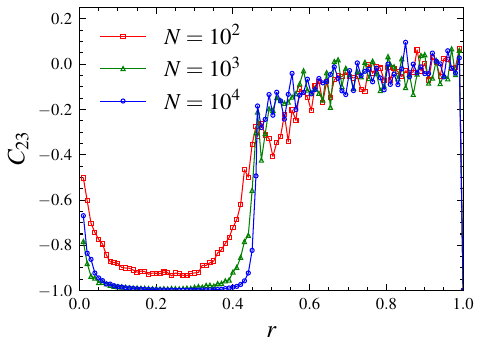}

    \caption{The variance $\sigma_{hh}$ 
    and correlation coefficients $C_{hh'}$ averaged over ten samples 
    with different trajectories. 
    $T=10^{-3}$, $t_0=100$, 
    $\tau=150$, $N=10^2, 10^3, 10^4$.
    } 
    \label{finitesize}
    \end{center}
\end{figure*}

Let us explain how to obtain sample trajectories by Monte Carlo simulations and compare them with the solutions of the derived population dynamics (\ref{PDeq}). As Monte Carlo simulations, we perform the following procedures:
\begin{enumerate}[(1)]
    \item[] 
    Initially, prepare a configuration $\bm{x}$ where the state at each site is a value taken randomly from $\{1,2,3\}$ with uniform distribution. That is, the probability of taking each state is $1/3$. 
    
    Next, repeat the steps from $1$ to $3$:
    \item Randomly choose a name $\ell\ (1 \leq \ell \leq N)$ of an agent with uniform distribution. That is, the probability of taking each name is $1/N$.
 
    \item With the probability $r$, 
  the chosen agent $\ell$ is disturbed, making the transition from state $x_\ell$ to $1$.
    
    \item If the chosen agent $\ell$ is not disturbed, choose either a positive or negative direction with probability $1/2$.
    \begin{enumerate}[(a)]
        \item If a positive direction is chosen and $x_\ell \neq 3$, make the transition from state $x_\ell$ to $x_\ell+1$ with the probability 
        \begin{equation}
            \frac{1}{2}\Bigg(1+ \tanh{\beta\Big(2(x_\ell-\frac{1}{N-1}\sum_{k \neq \ell}x_k)} + 1\Big)\Bigg).
        \end{equation}
        
        \item If a negative direction is chosen and $x_\ell \neq 1$, make the transition from state $x_\ell$ to $x_\ell-1$ with the probability 
        \begin{equation}
        \frac{1}{2}\Bigg(1+ \tanh\beta\Big(-2(x_\ell-\frac{1}{N-1}\sum_{k \neq \ell}x_k)+1\Big)\Bigg).
        \end{equation}
    \end{enumerate}
\end{enumerate}
As shown in Fig.\ \ref{dynamicsdata}, we have confirmed that Monte Carlo simulations have 
good agreement with the solution of the population dynamics (\ref{PDeq}). Note that for convenience in the simulations, we have replaced $\frac{1}{N-1}\sum_{k \neq \ell}x_k$ by $\frac{1}{N}\sum_{k}x_k$; the caused errors are the order of $1/N$, which is negligible, at least, near stationary states for sufficiently large $N$, except for the vicinity of $r=1, T=0$.

\subsection{Finite-size fluctuations}

Let us characterize the finite size fluctuations in the density of each state by variance $\sigma_{hh}$, covariance $\sigma_{hh'}$, and correlation coefficient $C_{hh'}$ for $h,h'\in\{1,2,3\}$, which are defined as
\begin{align}
    &\sigma_{hh'}:=\frac{N}{\tau-t_0}\sum_{t=t_0}^\tau\Delta_h(t)\Delta_{h'}(t),\\
    &\Delta_h(t):=\hat{Q}_h(t)-Q_h^*,\\
    &C_{hh'}:=\frac{\sigma_{hh'}}{\sqrt{\sigma_{hh}}\sqrt{\sigma_{h'h'}}},
\end{align} 
where $\hat{Q}_h(t)$ is a sample trajectory of density of state $h$ obtained from Monte Carlo simulation at time $t$. The time unit is one Monte Carlo step per site ($N$ steps). 
We set $t_0=100$ 
because the typical trajectories of $\hat{Q }_h(t)$ with $r\ge10^{-2}$ and $T=10^{-3}$ crosses the stationary solution much before $t=t_0=100$. 
Indeed, for $r\le10^{-3}$ and $T=10^{-3}$, 
the dynamics gets very slow and the typical trajectories do not cross
the stationary solutions by $t=t_0=100$.

As shown in Fig.\ \ref{finitesize}, 
we have performed Monte Carlo simulations with different systems sizes $N$. 
If we straightforwardly extrapolate the behaviors for larger system sizes, 
there seem to be singular behaviors in the three types of variance and the three types of correlation function in the large size limit near $r=r_0$ for $r<r_0$. 
In particular, it is notable to point out that three correlation coefficients in the large size limit seem to have perfect correlation, corresponding to $|C_{hh'}|=1$, 
for only $r<r_0$.

Additionally, the results obtained by Monte Carlo simulations suggest that  $C_{hh'}$ for $r<r_0$ tends to show U-shaped curves as the system size $N$ gets larger. It is notable that the bottoms of the U-shaped curves are close to $r=-5+2\sqrt{7}\simeq 0.291\cdots$, which is the point where $\mu_h$ changes its sign.
Nevertheless, more works with sufficiently large samples need to be done to obtain more convincing data, which remain to be found as another work in the future.

\end{document}